\documentclass[conference]{IEEEtran}

\IEEEoverridecommandlockouts
\usepackage{caption}

\usepackage{threeparttable}
\usepackage{tablefootnote}

\usepackage{algorithm}
\usepackage[noend]{algpseudocode}

\usepackage{varwidth}
\usepackage{amsmath}
\usepackage{xcolor} 
\usepackage{hyperref}

\usepackage{graphicx,epstopdf,algpseudocode,caption,url}   
\usepackage{multirow}  
\graphicspath{{./images}}

\usepackage{arydshln} 

\begin{document}

\title{Interaction in Metaverse: A Survey}  

\author{Hong Lin$ ^{1,2}$, Zirun Gan$ ^{3}$, Wensheng Gan$ ^{1,2*}$\thanks{\IEEEauthorrefmark{1}Corresponding author.}, Zhenlian Qi$ ^{4}$, Yuehua Wang$ ^{5}$, Philip S. Yu$ ^{6}$  \\ 
	\\
	$ ^{1} $Jinan University, Guangzhou 510632, China. 
	$ ^{2} $Pazhou Lab, Guangzhou 510330, China\\
        $ ^{3} $South China University of Technology, Guangzhou 510641, China\\ 
        $ ^{4} $Guangdong Eco-Engineering Polytechnic, Guangzhou 510520, China\\
        $ ^{5} $Tianjin University of Finance and Economics, Tianjin 300222, China\\    
        $ ^{6} $University of Illinois Chicago, Chicago, IL 60637, USA\\
        Email: wsgan001@gmail.com
}

\maketitle

\begin{abstract}
  Human-computer interaction (HCI) emerged with the birth of the computer and has been upgraded through decades of development. Metaverse has attracted a lot of interest with its immersive experience, and HCI is the entrance to the Metaverse for people. It is predictable that HCI will determine the immersion of the Metaverse. However, the technologies of HCI in Metaverse are not mature enough. There are many issues that we should address for HCI in the Metaverse. To this end, the purpose of this paper is to provide a systematic literature review on the key technologies and applications of HCI in the Metaverse. This paper is a comprehensive survey of HCI for the Metaverse, focusing on current technology, future directions, and challenges. First, we provide a brief overview of HCI in the Metaverse and their mutually exclusive relationships. Then, we summarize the evolution of HCI and its future characteristics in the Metaverse. Next, we envision and present the key technologies involved in HCI in the Metaverse. We also review recent case studies of HCI in the Metaverse. Finally, we highlight several challenges and future issues in this promising area.

\end{abstract}


\IEEEpeerreviewmaketitle

\section{Introduction}  \label{sec:introduction}

In the digital age \cite{gan2023web,sun2023internet}, Metaverse has recently not only become a trending word on social media but also a hot spot in the technology field \cite{sun2022big,sun2022metaverse}. Since Metaverse is still in the initial stage of exploration, there is no clear definition or consensus between the academic and industrial circles \cite{ng2022metaverse}. Overall, the definition of Metaverse can be roughly divided into three interpretations. First, it is the digitization of the material world \cite{mystakidis2022metaverse}. That is, on the basis of the digitalization of human vision and hearing realized by the Internet, Metaverse could realize a high degree of simulation of various sensory experiences. Second, it is a parallel world that allows humans to perform the same activities in the virtual space as in the real world, thus getting rid of the limitations of the physical space \cite{faraboschi2022virtual, huggett2020virtually}. Third, more generally speaking, it is a new type of Internet application and social formation, integrating multiple novel technologies and combining virtuality and reality. The virtual world and the real world will intersect and merge with each other on the levels of economic and social systems \cite{sun2022metaverse, chen2022metaverse}. Metaverse relies on artificial intelligence, human-computer interaction, digital twins, and other technologies as the underlying architecture. Technological innovations in many aspects have made Metaverse considered a new form of future Internet development. At present, there have been relevant studies exploring the application of Metaverse in multiple fields, including medicine \cite{wu2022scoping,kovacev2022metaverse}, education \cite{lin2022metaverse}, manufacturing \cite{yang2022parallel}, etc. One of the characteristics of the Metaverse is decentralization, which aims to build a new operating order for the Internet. It realizes distributed storage data and distributed computing data based on blockchain technology, ensuring personal privacy security and authenticity authentication \cite{ning2023survey}. With the Metaverse, enterprises can connect with customers, interact with them, and motivate them to create new exchanges of value and streams of revenue. For the public, the immersive experience is the most attractive point of Metaverse, and its ultimate goal is to make users feel as if they are in the real world. This not only requires high-precision and low-latency effects but also simulates a variety of human sensations. The above requirements are inseparable from human-computer interaction (HCI) \cite{lazar2017research}, which is the key technology to realizing immersive experience.

HCI is an information exchange process between humans and computers to complete a certain task \cite{lazar2017research}. It focuses on the three parts of human, machine, and interaction, aiming to efficiently realize the conversation between humans and computers through input and output devices. On the one hand, the machine provides people with a large amount of relevant information and prompts for instructions through the output device. On the other hand, people transmit relevant information to the machine through the input device, answer questions, and prompt instructions. Therefore, it involves not only various computer science disciplines, including image processing \cite{su2015virtual}, computer vision \cite{suma2019computer}, etc., but also multiple fields in the humanities, such as ergonomics \cite{eason1991ergonomic}, human factors engineering \cite{bannon1995human} and cognitive psychology \cite{carroll1997human}, among others. As the name implies, HCI has been studied since the birth of computers. After decades of development, HCI has evolved from the original manual operation to the ``Command-line User Interface (CUI)" and ``Graphical User Interface (GUI)". Although the efficiency of interaction has improved, it requires users to learn relevant instructions to interact with the computer, resulting in a lack of immersion. A GUI based on virtual reality and augmented reality is a current solution that provides users with visual immersion through a 3D user interface \cite{bowman2006new}. Users select a certain point or area in the image through the controller to complete operations such as zooming and dragging the virtual object. While technologies such as voice and gesture recognition are now beginning to be combined to reduce reliance on controllers, such solutions still cannot completely get rid of physical devices that would reduce immersion.

Overall, device-dependent interactions will be gradually eliminated, and new forms of interactions and business models across multiple applications will emerge. Examples of early-stage solutions include virtual collaboration, navigation apps, social media, and fungible and non-fungible tokens. Besides, a new solution, namely ``Natural User Interface (NUI)"\footnote{\url{https://wikipedia.org/wiki/Natural_user_interface}}, is proposed. Its goal is that people can immerse themselves in the interaction with computers just like their daily communication via voice and hand recognition \cite{falcao2015evaluation}. Users do not need to use physical devices such as a mouse and keyboard for manipulation. In addition, the real-time multilingual capabilities of conversational AI will break down language barriers and facilitate interaction in the globally connected Metaverse.

\textbf{Contributions}: To fill this gap, this paper aims to provide a systematic literature overview of HCI in Metaverse. The contributions to this review are as follows:

\begin{itemize}
    \item We analyze the relationship between HCI and the metaverse and conclude that the development of both is mutually reinforcing.
    
    \item We review the history of HCI and divide it into different eras, and envision the future of HCI, including the characteristics of the interaction and related technologies.

    \item We list the more popular and mature HCI products currently available and summarize their interaction paradigms and characteristics.
    
    \item We analyze in detail the challenges and issues encountered by HCI, including technological constraints, lack of accuracy, and resource consumption.
\end{itemize}

\textbf{Roadmap}: We analyzed the relationship between Metaverse and HCI in Section \ref{sec:Relationship}, and listed what effect Metaverse needs HCI to achieve in Section \ref{sec:Effect}. Then, we introduce the key technologies related to HCI in Section \ref{sec:Technologies}, and summarize the most advanced industry case applications in Section \ref{sec:Cases}. Then, in Section \ref{sec:Challenges}, we evaluate the challenges and issues faced by HCI in the Metaverse. Section \ref{sec:Conclusion} concludes this paper with a discussion of potential future research. Table \ref{Symbols} describes some basic symbols in this article.

\begin{table}[ht]
    \small
    \centering
    \caption{Summary of symbols and their explanations}
    \label{Symbols}
    \begin{tabular}{|c|c|}
    \hline
\textbf{Symbol} & \textbf{Definition}               \\ \hline
AI              & Artificial intelligence           \\ \hline
AR              & Augmented reality                 \\ \hline
CAD             & Computer-aided design             \\ \hline
CPS             & Cyber-physical systems            \\ \hline
EEG             & Electroencephalogram              \\ \hline
HCI             & Human-computer interaction        \\ \hline
MR              & Mixed reality                     \\ \hline
NBA             & National basketball association   \\ \hline
NLP             & Natural language processing       \\ \hline
UFC             & Ultimate fighting championship    \\ \hline
VR              & Virtual reality                   \\ \hline
XR              & Extended reality or cross reality \\ \hline
5G              & 5th generation mobile network     \\ \hline
6G              & 6th generation mobile network     \\ \hline
\end{tabular}
\end{table}

\section{Relationship Between Metaverse and HCI}
\label{sec:Relationship}

The foundation of the Metaverse is a series of cutting-edge technologies, such as artificial intelligence, blockchain, Web3, etc., that build a world where virtual and reality are interconnected and interoperable. What can really derive and magnify value lies in interaction, that is, various interactions generated by various applications. The greater and stronger the dependence on interaction, the greater its corresponding commercial value. HCI is the bridge connecting the real world and the virtual world of the Metaverse, and it is also the entrance for people to enter the Metaverse, which determines their immersion in the Metaverse. Admittedly, the virtual world sounds cool, but it still cannot be realized without the support of hardware carriers. VR and AR, as important hardware carriers of the Metaverse, have always been regarded as ``the ladder to provide immersive virtual reality experiences for Metaverse users". In addition, according to a recent report \cite{aaron2022metaverse}, Metaverse has received investment from many companies since its rise in the past two years. As an emerging concept, why can Metaverse attract attention and investment capital so quickly? Largely because VR and AR have relatively mature products, including headsets, smart glasses, and tactile gloves. After all, people will not easily pay for a fantasy idea. On the contrary, they have seen the embryonic form of the Metaverse through VR/AR technology.

\begin{figure}[ht]
    \centering
    \includegraphics[trim=5 0 0 0,clip,scale=0.32]{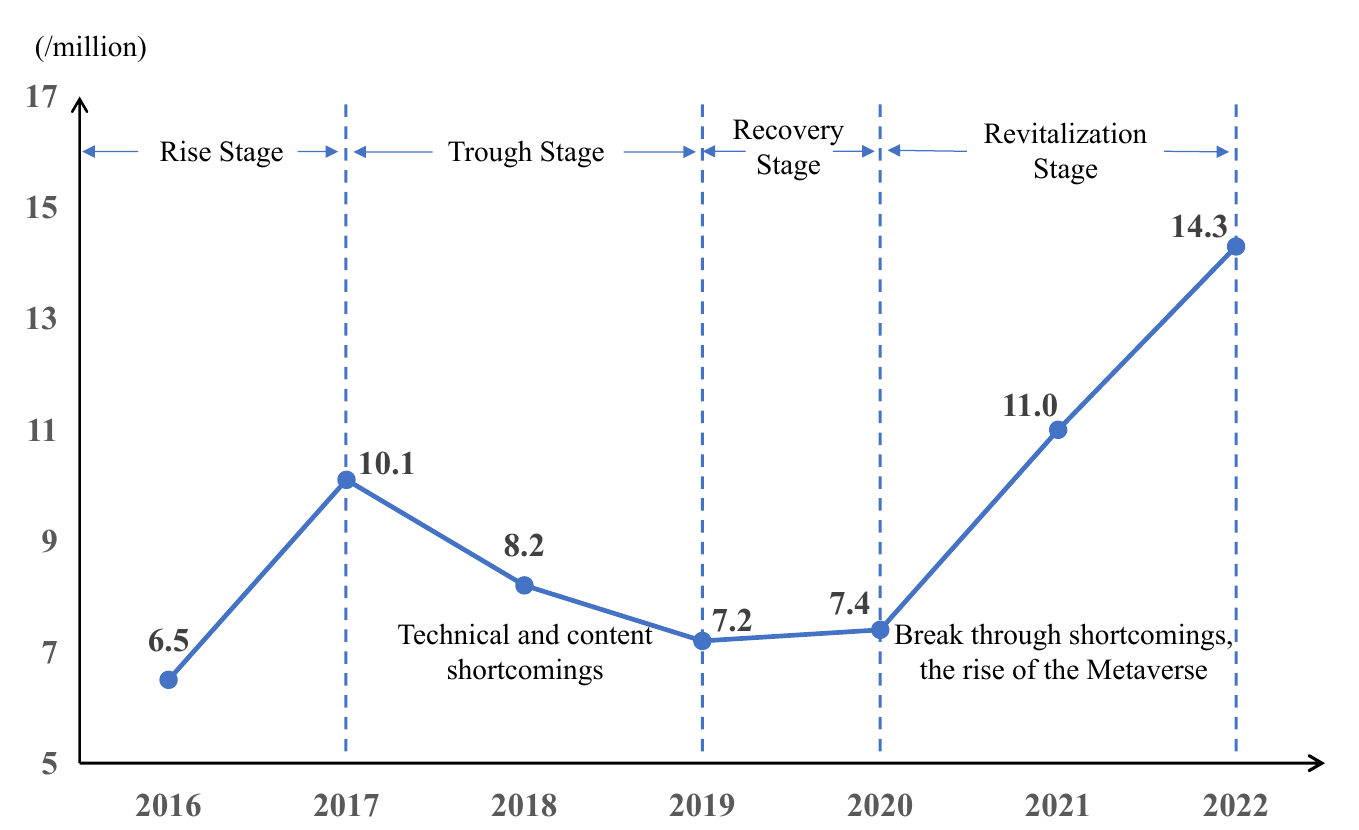}
    \caption{Global shipments of VR/AR devices at different stages}
    \label{fig:shipments}
\end{figure}

From another perspective, the Metaverse is equally important to the development of HCI. Take the VR/AR industry as an example. Since 2012, a large number of VR/AR devices have emerged and entered the market. They were once considered the next-generation general-purpose computing platform to replace smartphones. However, in 2017, due to the fact that the business model was still unclear and the bottlenecks in network, hardware, and content had not been broken through, people gradually revised their expectations for the ``bright future" of VR/AR. The capital market has tightened investment, and the progress of the VR/AR industry has slowed down and entered a trough. Subsequently, the VR/AR industry began to focus on overcoming the core shortcomings of hardware and content ecology, and the market has gradually recovered since 2019. Coinciding with the rise of the Metaverse and 5G, global shipments of VR/AR equipment have risen sharply since 2020. Based on the statistical data \cite{sara2022ar}, Fig. \ref{fig:shipments} shows global shipments of AR/VR devices at different stages. The needs of the Metaverse also provide directions for the research of VR/AR and the entire HCI. We will discuss the goals that HCI in the Metaverse may achieve in the future.

\section{History of HCI} \label{sec:Effect}

Since people need to deal with computers, HCI appeared with the birth of computers. The earliest computers were huge monsters composed of electronic tubes. People entered commands based on punched cards and then represented the binary (0 or 1) of the execution results through the electronic tube lights (light or dark). This kind of ``manual operation" interaction is very troublesome in information input and reading, inefficient and error-prone, and requires high professionalism. After the 1960s, command-line computers with electronic screens appeared, which input commands based on the keyboard and then output the execution results to the screen in the form of character text. This ``Command-line User Interface (CUI)" interaction greatly improves the efficiency of interaction, until modern computers still retain this method (e.g., the DOS command line of Windows OS). But it still requires high professionalism, which means memorizing a lot of commands. In the 1980s, Apple Inc. released the computer Lisa, which pioneered elements such as windows and icons to represent corresponding operations, marking the birth of the ``Graphical User Interface (GUI)''. Users input instructions based on the mouse and keyboard, and the interactive content also becomes multimedia (text, graphics, and images). Later, with the appearance of sound cards, loudspeakers, etc., the interactive content added audio and video. GUI interaction enables users to understand the meaning of the interface more intuitively; the content is richer and more interesting, and the operation is more convenient. Moreover, it is no longer limited to the use of professionals but is available to the public. Until now, it has still been the mainstream human-computer interaction method, but it requires users to master the pre-set operation rules (e.g., the meaning of the icon, and the shortcut key of the keyboard). The popularity of touchscreen-based smartphones has brought people into the Post-PC era\footnote{\url{https://en.wikipedia.org/wiki/Post-PC_era}}. With the application of sensors and artificial intelligence, voice interaction and gesture interaction bring a non-touch interaction experience for people, trying to achieve ``natural interaction interface (NUI)", although they are still not mature enough. Finally, as shown in Table \ref{tab:compare}, we compared HCI in different eras in several aspects.

\begin{table*}[ht]
    \caption{Comparison of HCI in different eras}
    \centering
    \includegraphics[trim=0 0 0 0,clip,scale=0.31]{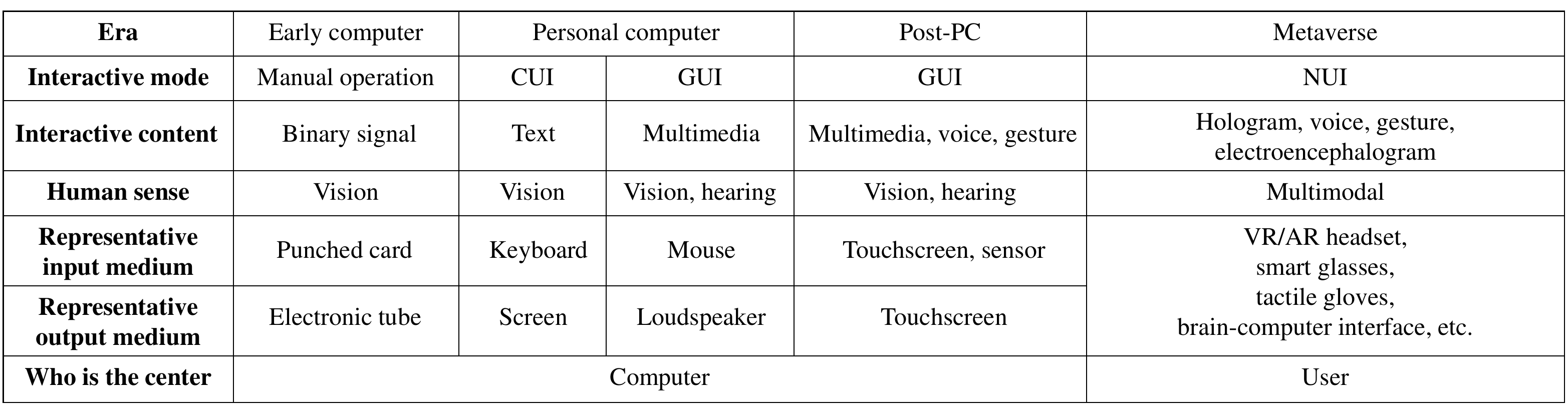}
    \label{tab:compare}
\end{table*}

\section{Future prospect of HCI in Metaverse}

After reviewing the development of HCI, we can summarize some rules. The operation of HCI is gradually becoming easier, and as the content becomes richer, its audience expands from professionals to the general public. In general, the relationship between humans and machines has become closer, moving toward NUI. Based on these trends, we envision the future of HCI in the following.

\subsection{User-centered interaction}

The development of artificial intelligence is changing the traditional HCI model of human-machine relationships. The concept of user-centered interaction is also being newly articulated, rather than focusing only on more fashionable design, easier operation, etc. For example, instead of passively receiving user input and executing instructions, future machines can always capture and understand the user's state (e.g., emotions, intentions, etc.) and quickly respond in a rational manner. Such reactions are intelligently deduced based on the user's personal characteristics (e.g., identity, hobbies, consumption behavior, etc.) or environment. In addition, the user is able to predict the behavior of the machine with the help of a model and trust the machine's decisions. In short, the device is not merely a tool but more like a companion,  capable of providing solutions when the user needs assistance \cite{xu2021user}.

\subsection{Multimodal interaction} 

Modality is a biological concept that describes the channels through which organisms receive information from their senses and experiences. For example, humans have senses such as sight, hearing, touch, taste, and smell. In the field of HCI, researchers have developed single-modality interactive interfaces based on different modalities, such as graphical user interfaces and speech user interfaces. However, they cannot fully meet the needs of interaction, because, in reality, human-human interaction is essentially a combination of multiple modalities. For example, we can recognize each other's joyful mood from hearing (e.g., laughter), vision (e.g., smile), and gesture (e.g., body language). Multimodal interaction refers to the interaction of people communicating with computers through voice, body language, and other information carriers (text, pictures, audio, video, etc.). It is based on multimodal machine learning and integrates voice interaction, eye tracking, haptic interaction, and somatic interaction, aiming to fully simulate human-human interaction or human-environment interaction. In the future Metaverse, it will establish Natural User Interfaces (NUI) to achieve efficient and reliable information transfer between humans and machines.

\subsection{Biometrics-based interaction}

With the spread of the Internet, humans are now moving from the physical world into the digital world, and this "Great Migration" will be completed in the Metaverse era. As in real life, the question "Who am I?" will still exist in digital life because it is the first step of HCI. Moreover, as more information is shared between humans and machines, more security will be required for identification. Password-based identification is becoming increasingly vulnerable, and the future of reliable identification comes from the biometrics of each individual \cite{gupta2023survey}. Due to the uniqueness of biometric features, it will be the most stringent identification for humans entering the digital world. Biometric identifiers include physiological characteristics and behavioral characteristics. There are other, more secure and stable options (e.g., DNA \cite{zahid2019biometric}) waiting to be promoted and applied. In addition, since the activity space of HCI in the future is no longer limited to the screen, the recognition of behavioral features will also become a solution. It includes gait \cite{wan2018survey}, signature \cite{poddar2020offline}, etc.

Based on the above characteristics, we propose the steps of HCI in the Metaverse. Although different products may bring different interaction paradigms in the future, they are all inseparable from these steps:

\begin{itemize}
    \item \textbf{Identification}: Verify the user's identity based on biometrics to ensure that the corresponding interaction effects are provided based on the user personas. 

    \item  \textbf{Signal acquisition}: Collecting signals from different parts of the body with sensors.

    \item  \textbf{Signal integration}: Integrate signals and remove redundant signals to form a global signal.

    \item  \textbf{Process}: Infer user interaction intent based on integrated global signals.

    \item  \textbf{Task execution}: Sends the user's interactive commands to the application for execution.

    \item  \textbf{Feedback}: Provide the execution result to the user, indicating the completion of the interactive command.
\end{itemize}

\begin{figure}[ht]
    \centering
    \includegraphics[trim=5 0 0 0,clip,scale=0.42]{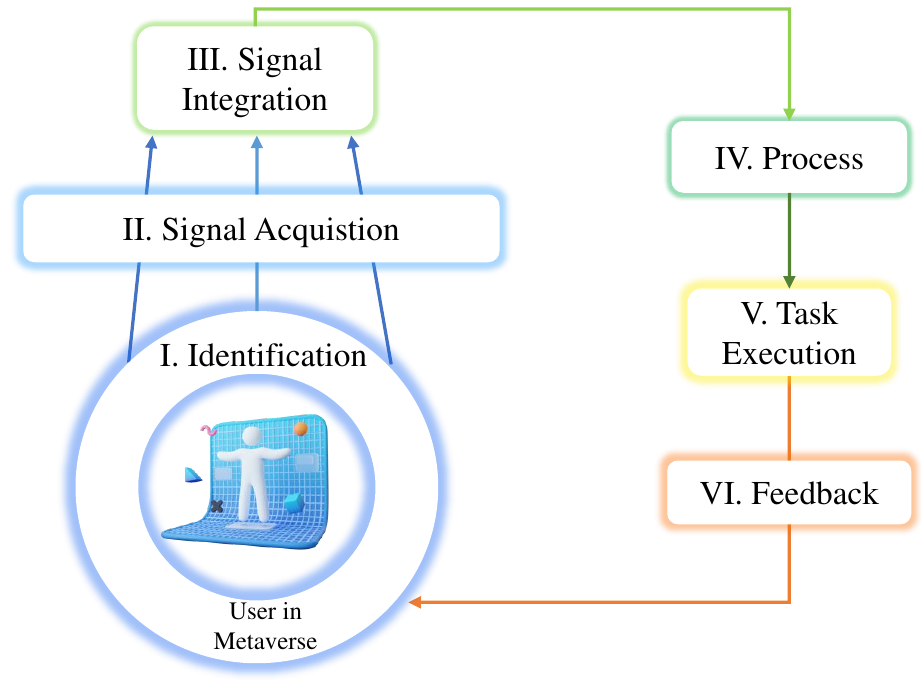}
    \caption{The steps of HCI in the Metaverse}
    \label{fig:steps}
\end{figure}

\section{HCI Technologies Required in Metaverse}
\label{sec:Technologies}

\begin{figure}[ht]
    \centering
    \includegraphics[trim=5 0 0 0,clip,scale=0.3]{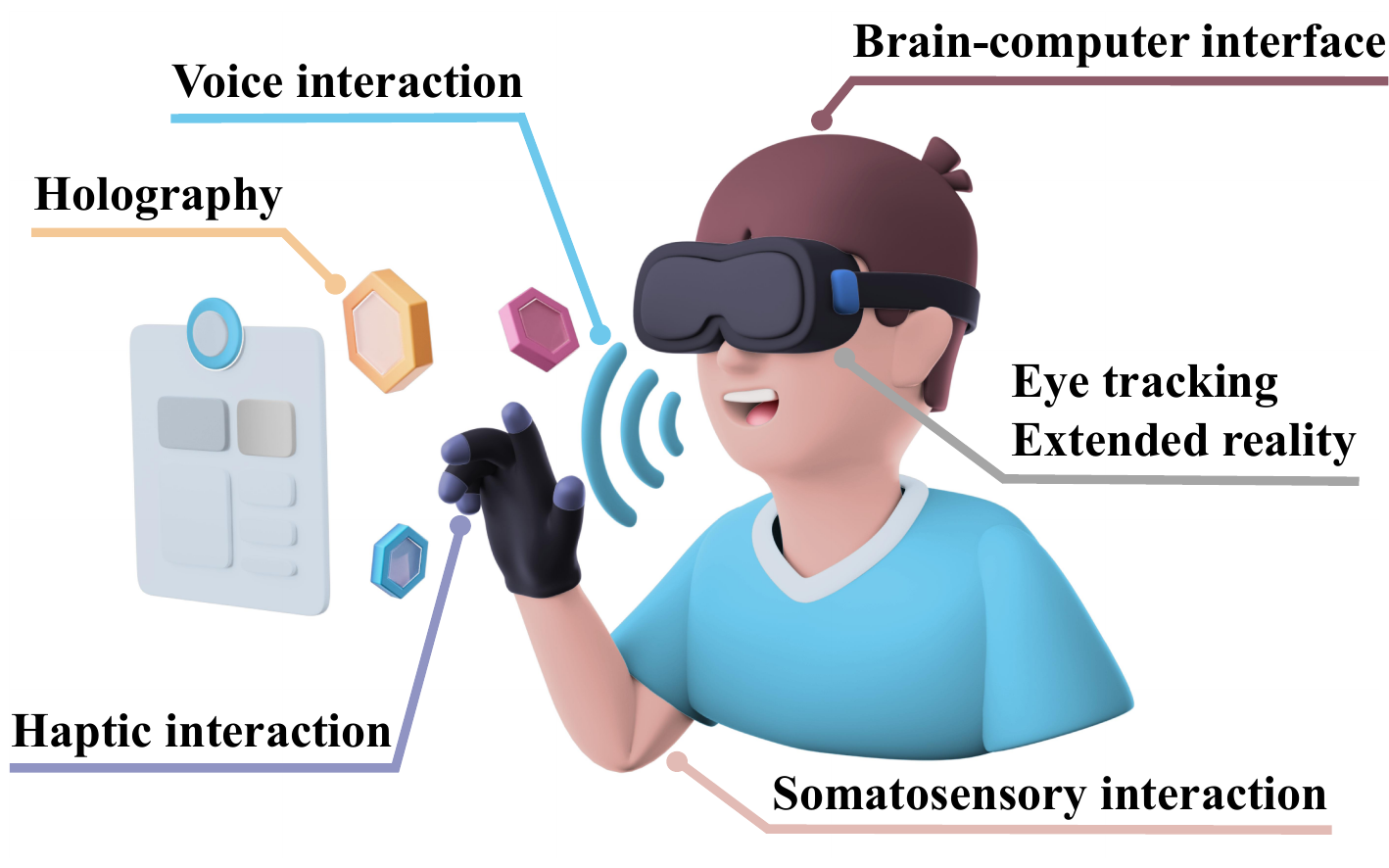}
    \caption{HCI technologies required by users in Metaverse}
    \label{fig:technology}
\end{figure}

In this section, we introduce the HCI technologies that users need in Metaverse and the current state of research. As shown in Fig. \ref{fig:technology}, they are based on the various sensory parts of the user and cooperate to bring a rich interactive experience.

\subsection{Sensory interaction}

\textbf{Voice interaction} is a common mode of modern HCI. Users’ words are processed by automatic speech recognition and natural language processing (NLP) and transformed into instructions that can be recognized by the computer. Some devices, so-called smart voice assistants, provide a text-to-speech system to convert normal language text into speech feedback\footnote{https://en.wikipedia.org/wiki/Speech\_synthesis}. Recently, benefiting from breakthroughs in NLP enabled by machine learning and deep learning, intelligent voice assistants have become widely popular in devices such as mobile phones, intelligent housing systems \cite{hasan2017innovative}, and cars. Users can operate smart devices by giving short voice commands without having to input words or press any buttons. Matthias conducts a survey and demonstrates that voice interaction reduces distractions and operational difficulties with smart devices in cars, thereby preventing accidents \cite{peissner2011can}. Meanwhile, voice commands also contain more information, such as voiceprints and emotions, which can be applied to user authentication and emotion detection.

\textbf{Eye tracking} is a technology that measures users’ eye movements with an infrared camera \cite{poole2006eye}. Just and Carpenter assume that the eye movements of people can reveal their attention and cognitive processes \cite{just1976eye}. HCI researchers have studied for many years to find out the “areas of interest” through eye movement and the change of the facial area near the eye. They try to project the continuous recordings of eye movement into a series of information. For VR or AR, these recordings are often projected into the user's perspective information, so the renderer can focus on what the user is looking at and reduce the rendering of other areas. This will improve the performance of VR or AR devices and prevent users from experiencing vertigo. Other studies have shown that a gaze-control multimedia interface with eye tracking can reduce error rates in immersive environments when compared to a keyboard \cite{sidorakis2015binocular}. Other health-monitoring devices, such as onboard fatigue sensors, monitor the user's eye movement to identify whether drivers are tired or distracted \cite{palinko2010estimating,trosterer2014eye}.

\textbf{Somatosensory interaction}, in which commands are expressed by actions (mainly gestures and limb movements), has the great advantage of being eyes-free and device-free. However, it is hard to widely use in interaction systems, even though somatosensory interaction has been invented for many years and deep vision technology has greatly improved gesture recognition in recent years. There are two main reasons. First, in the user's action instruction, the action with interactive intention is always mixed with the action without interactive intention. The gestures or actions of users are often mixed with redundant actions, which greatly improves the difficulty of recognition. Second, the difficulty of mapping action to instruction for users is high. Even the action instructions made by the same user are not exactly the same each time, so it is difficult for the system to establish a robust action-instruction mapping relationship.

\textbf{Haptic interaction} plays an important role in the interaction of Metaverse. Adams \textit{et al.} \cite{adams1999stable} describe haptic interaction as a sensual link between humans and machines. The largest organ of the human body is the skin, and humans acquire information from the external world through it \cite{dahiya2009tactile}. Researchers have extensively studied how to implement touch recognition and force feedback in interactive devices. Smart gloves are the most common haptic interactive device and have been used in VR/AR, augmentative and alternative communication, and rehabilitation fields. Through force-and-flex sensors, smart gloves detect the force and bending angle of fingers to recognize the motion of fingers and palms \cite{ozioko2022smart}. They then transfer the motion into operations and send the operations to the machine. Also, some devices can generate feedback such as vibration, heat, shape-changing, and friction.

\subsection{Brain-computer interface}

Brain-computer interface  (BCI) bridges the human neural world with the outside physical world by converting brain signals into commands that can be recognized by computer devices. The process mainly includes signal acquisition, signal processing, device control, and signal feedback \cite{camargo2021brain}. According to the access method, BCI can be divided into invasive and non-invasive. Invasive BCI refers to the implantation of hardware devices such as chips and sensors into the cranial cavity or electrodes into the cerebral cortex through surgery, etc. Its advantage is that the quality of neural signals obtained is high, but it may lead to immune reactions and damage to the brain. Non-invasive BCI refers to recording and interpreting brain information without invading the brain through a wearable device attached to the scalp.
It is safer, but the acquired neural signals will be attenuated and interfered with due to the blockage of the cranium \cite{salahuddin2021signal}.

Most of the research directions for HCI methods allow users to interact with computers with the help of their own sensory systems and limbs, enhancing the ease of interaction. However, BCI will bring a disruptive change in which users only need to transmit commands and receive information through their brains, without the involvement of other parts of the body. Compared with other HCI methods, the breakthrough of BCI will be a milestone for the Metaverse. First, users will have more freedom of movement in the Metaverse, unrestricted by pre-defined interaction commands. In addition, multi-sensory interaction will become possible through the bidirectional transmission of brain signals, bringing a richer experience for users. Furthermore, the BCI will enhance the immersion of the Metaverse. The ``Rubber Hand Illusion" experiment\footnote{\url{https://wikipedia.org/wiki/Body\_transfer\_illusion\#Rubber\_hand\_illusion}} proves that the human brain is inclusive, and it can dynamically adjust the sense of boundary between humans and the outside world. This means that when the brain is connected to an external device through BCI, it will have the possibility to accept and assign new neurons to the external device. Thus, BCI can achieve a deeply immersive experience by tricking the brain's sense of boundaries with the external world.

\subsection{Extended reality}

\textbf{Virtual reality}. VR is a technology that can build an immersive three-dimensional world in users’ eyes \cite{huang2018augmented}. Most VR systems rely on VR head-mounted displays. Inside the head-mounted display, there are convex lenses to enlarge the range of images seen by the user’s eyes and generate a visual field of 180-240 degrees for users. Then, the gyroscope of the head-mounted display calculates the angles that users have moved and informs the image engine to update the images in time. The real-time changing images make users feel that they are in a virtual immersive world. Additionally, the images perceived by both eyes are different, creating a pronounced three-dimensional vertical sensation. To build an immersive virtual world, we also need super screens with high resolution and high refresh rates. According to \cite{mangiante2017vr}, 8K quality per eye is required since the head-mounted display is a few centimeters from the eyes and the visual effect of a 4K display is the same as 240p in the TV screen. As for the refresh rate, a high refresh rate helps the VR system cheat users’ eyes and the VR with a low refresh rate might cause dizziness and nausea in users. Most VR systems are designed for completely immersive scenarios such as video games and virtual meetings. There is also research to apply VR to virtual tourism, art, industry, and education. Because of technical restrictions, it is still a long way to achieve these goals. In recent years, there are over 230 companies such as Microsoft, Apple, and Meta, have produced VR or performed research related to VR \cite{park2019literature}. VR is expected to popularize in millions of families within 10 years and become a new mainstream HCI way.

\textbf{Augmented reality}. AR integrates virtual objects and the natural world \cite{huang2018augmented}. It projects the virtual object into the physical world through video synthesis or optical projection. Most AR system creates a mapping between the real world and virtual objects by computer vision. The camera of the AR system captures the visual information and determines the baseplate or some projection points of virtual objects. Then the image engine zooms in and out the size of virtual objects and renders the graphics or 3D models on the screen as if they were attached to real objects. Unlike VR, AR is a technology that enhances the real world instead of building an immersive virtual world. Users can always see the real world, so the immersion is not the total \cite{elmqaddem2019augmented}. AR makes users feel that virtual objects exist in the real world and can interact with virtual objects. Compared to VR, AR is more suitable for the scenarios that users need to see the physical world. In these scenarios, AR is an auxiliary way to display real-time information according to the physical world. The helmet-mounted display system of fighter pilots is the best example of AR. The other application is to use AR for presenting information in education \cite{elmqaddem2019augmented}, medical treatment, tourism, entertainment, and industry \cite{azuma1997survey}.

\subsection{Holography}

While extended reality offers a solution for immersive visual experiences, it comes with compromises. One such compromise is the comfort of the headset. Additionally, most AR/VR requires that the user's eyes receive slightly different images separately and adjust to a fixed focal length to create the illusion of 3D \cite{the2022holography}. However, the user is limited in observing the image from different perspectives. Moreover, this stereoscopic imaging method can have physiological side effects, such as dizziness and nausea \cite{tomas2022way}, a phenomenon known as Vergence-accommodation conflict (VAC) \footnote{\url{https://en.wikipedia.org/wiki/Vergence-accommodation_conflict}}. Holography is a way to address these shortcomings, which is regarded as the ultimate goal of the Metaverse in terms of visual experience. The effect of holography is not new, as it is widely seen in science fiction movies (e.g., Star Wars, Avatar). Images are displayed accurately in the physical world in 3D, and a person can see different perspectives of the holograms directly through the naked eye by changing the perspective at will. It does not require a headset and has no VAC, providing a fully immersive visual interaction.

The original holography is optical holography, but it suffers from the fact that the displayed hologram is static, and the optical conditions are harsh. Computer-generated holography solves these limitations, and the display is no longer limited to physical 3D objects but can also be digital 3D models \cite{liangcai2022progress}. Although some products are currently advertised as having holograms, they are not actually true holography in terms of technical implementation. At present, holography in the true sense is not yet commercially available \cite{chang2020toward}. With the help of an optical illusion called Pepper's ghost \footnote{\url{https://en.wikipedia.org/wiki/Pepper's_ghost}}, a hologram-like effect can be achieved by projecting the image onto a special medium (not air). The Metaverse puts extremely high demands on holography, which has to break through many challenges. The latest holography research is mainly based on deep learning in terms of algorithms \cite{shi2021towards, zhang2022progress}, and accompanied by hardware innovation \cite{jiang2019improve, li2020scalable, an2020slim}.

\subsection{Artificial intelligence}
Interactive text plays an important part in HCI. For traditional interactive content, developers prepare text in advance and update text in the program periodically. However, the production of content is hard to keep pace with consumption, and it's easy for users to get bored with existing content. After the success in AI, researchers released many NLP models \cite{kenton2019bert, vaswani2017attention, radford2018improving} (e.g., Transformer, BERT, GPT) and AI generative text is expected to solve the problem. AI generative text is the text content generated by AI. It is widely used in the commercialization of structured generative text with a relatively fixed form. For example, AX Semantics can generate product descriptions, news, and financial reports without any code. However, generating long creative text content, such as fiction, blogs, and poetry, is much more challenging. With the development of communication and Internet technology, the demand for generative interactive text is also growing rapidly. ChatGPT, a chatbot developed by OpenAI, was introduced on Nov 22, 2022, and is regarded as the world's best chatbot at the time. ChatGPT is popularized in the world because users found that they can get a detailed and human-like reply from it. ChatGPT's ability is far more than people expected and it has been applied to different areas \cite{jiao2023chatgpt, aydin2022openai}. 

\subsection{Edge computing}

Even the current advanced Metaverse architecture still relies on a centralized cloud-based design for virtual physical simulation and graphics rendering computations \cite{dhelim2022edge}, and this way of transferring heavy computational operations makes the HCI experience unsatisfactory, e.g., high latency and low-quality visualization. Edge computing is an architecture that enables data storage and processing closer to the network edge of the data generated by HCI. Although it does not have the same computing resources as cloud computing, it can effectively deliver a high-quality and immersive interactive HCI experience due to the geographic proximity to the device \cite{zhang2022uav}. In addition, because data is stored locally rather than remotely in the cloud, edge computing reduces the risk posed when cloud servers are attacked \cite{cao2020overview}.

\section{Industry Case Studies of HCI in Metaverse}
\label{sec:Cases}

We are looking forward to the development of HCI in the Metaverse, which will determine the future of the Metaverse. As shown in Table \ref{tab:company}, there are already several groups focused on developing advanced HCI products and defining new interaction paradigms, including ``Headset + Controllers'' for VR products, ``Headset + Voice + Gesture'' for AR products, and ``BCI devices + EEG signals'' for BCI products.

\begin{table*}[ht]
    \caption{Several representative and productive cases}
    \centering
    \includegraphics[trim=0 0 0 0,clip,scale=0.38]{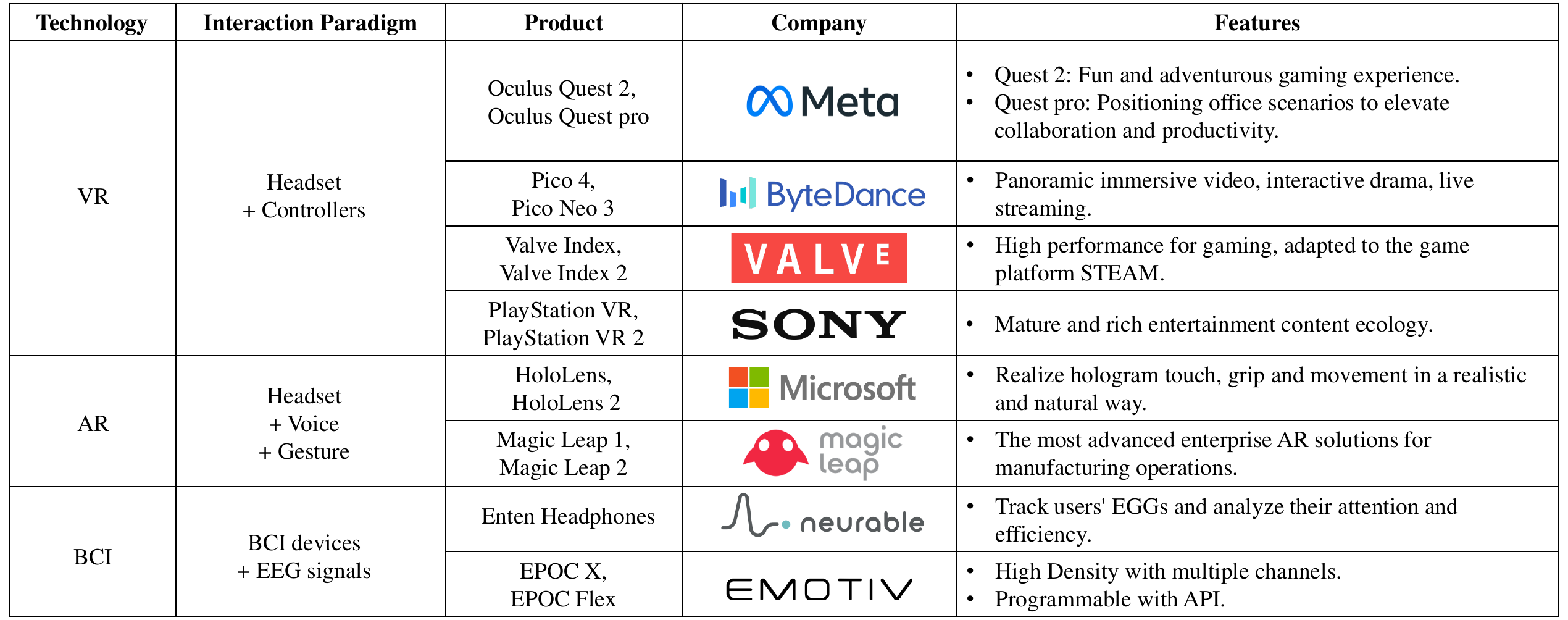}
    \label{tab:company}
\end{table*}

\subsection{Microsoft HoloLens}

Microsoft HoloLens\footnote{\url{https://www.microsoft.com/en-us/hololens}} is currently one of the hottest wearable devices and has been released in two generations. It can recreate a virtual space resembling the real environment and generate digital content to bridge the gap between the virtual and reality. Additionally, it provides users with a rich solution for interaction with virtual digital content through ``Gaze", ``Gesture" and ``Voice"\footnote{\url{https://learn.microsoft.com/en-us/windows/mixed-reality/design/system-gesture}}. Gaze is used for interaction target selection, which is essential, while Gesture and Voice are used to interact with the content. They are designed to touch, grasp, and move holograms in a natural way.

\begin{itemize}
    \item \textbf{Gaze}: The initial approach of Gaze is head-gaze, similar to dragging the mouse by hand, which moves the cursor according to the rotation of the head, possibly resulting in user fatigue. HoloLens 2 additionally implements eye-gaze based on eye tracking, which only requires sight through the eyes with little effort. It currently needs to be more accurate in head-gaze in selecting tiny targets.

    \item \textbf{Gesture}: Gesture refers to direct interaction with hands, including grasping and pressing, with the ultimate goal of allowing users to manipulate objects as naturally as they would in the real world. Hololens has several simple predefined gestures that need to be learned by users, while Hololens 2 uses instinctual gestures and no longer requires users to remember gesture commands.

    \item \textbf{Voice}: Voice is a solution for users who are not convenient to use gestures. Based on voice interaction, Hololens has introduced a voice mode called ``see it, say it", where the text of labels on the buttons can be activated as language commands while being able to converse with the digital agent.
\end{itemize}

\subsection{Meta Quest Pro}
 Meta Quest Pro\footnote{\url{https://www.meta.com/quest/quest-pro/}} is an all-in-one VR helmet released by Meta (formally Facebook)\footnote{\url{https://about.meta.com/}} in October 2022. It is a VR helmet that works independently and connects to a PC. As the annual flagship product of 2022, Quest Pro has received great attention since its launch. Quest Pro is equipped with Qualcomm Snapdragon XR2+ to provide high-performance computing power, AI, and 5G network access. Moreover, Quest Pro uses two new LCD monitors with high resolution and high refresh rate to effectively reduce users' dizziness. In terms of interactions, Quest Pro implements the avatar of users, motion tracking, and synthesis of virtual and real.
 
 \begin{itemize}
     \item \textbf{Avatar}: Users can join a virtual world and create their avatars corresponding with themselves. Avatars provide the volume of bodies and a real sense of scale in the virtual world with real volumes associated with the virtual world in which users are\footnote{\url{https://developer.oculus.com/resources/bp-avatars/}}. Quest Pro has five facial sensors that capture the motions of the eyes and face and create a variety of expressions that can be displayed on the avatar. 
     
     \item \textbf{Motion tracking}: The new motion controller, Oculus Touch Pro, has onboard gesture tracking. Each controller has three built-in sensors that track its position in 3D space without the help of a helmet, giving users 360-degree tracking. With the help of controllers, avatars can perform the same actions as the user, manipulate objects in the virtual space, and interact with other users. 

     \item \textbf{Synthesis of the virtual and the real}: With sensors and high-resolution external cameras, Quest Pro blends the physical world and virtual space. The synthetic scene presented through the screen is almost as real as the physical world. The improved experience allows users to transport themselves into work and entertainment combining the physical world and virtual space.   
 \end{itemize}

\subsection{Emotiv EPOC}
Emotiv is the market leader in brain-computer interface and has released a variety of products. Its flagship product is the EPOC series, which includes the EPOC X\footnote{\url{https://www.emotiv.com/epoc-x/}} and EPOC Flex\footnote{\url{https://www.emotiv.com/epoc-flex/}}. The EPOC series features multiple electrodes (14 or 32) for more accurate acquisition of EEG signals. The light weight makes it comfortable to wear, and it also supports wireless Bluetooth connectivity. In addition, Emotiv provides an application programming interface, Cortex API, to acquire the user's EEG signals for training and recognition of mental instructions. The API is based on JSON and WebSockets\footnote{\url{https://www.emotiv.com/developer/}}, making it easy to access from a variety of programming languages and platforms. Using this API, users can develop third-party applications, making the application scenarios of EPOC series very diverse, such as controlling a miniature car and a virtual reality Wheelchair.

\section{Challenges and Issues} \label{sec:Challenges}
\subsection{Technical constraints}

The current technologies are far from realizing the ideal Metaverse described in the concept. As a medium between the real world and the virtual world of the Metaverse, the interactive device of HCI needs to be convenient, flexible, portable, and can better immerse users in the virtual world. Nowadays, accessing the virtual world primarily depends on XR devices with high immersion. However, current virtual implementation technology faces challenges in terms of miniaturization, portability, and affordability. Prolonged use of XR devices can also lead to discomfort \cite{sun2022metaverse}. Somatosensory interaction and haptic interaction, like other existing interactive technologies, also face difficulties in sensing and identification technology, which makes them difficult to popularize. At the same time, the pursuit of lighter and more comfortable devices limits computing power and battery life. Although BCI can enhance the ease of interaction effectively, it faces a high risk of surgery and human tissue rejection \cite{king2022risks}. Non-invasive BCI can partially mitigate the safety risks associated with complex surgery, but it faces challenges due to weak signal collection and limited dissemination. The interactive mode is a higher-level problem for HCI in the Metaverse. The machine needs to change its identity from a passive responder to an active server. However, due to the diversity of user terminals, the development of terminal application functions has become another challenge for HCI. As a virtual world that all users can access anytime and anywhere, how to control target objects, express abstract commands, and input the content in the virtual world are also key technologies for HCI that need to be broken through in the Metaverse. It is difficult to maintain and switch the interactive state of multiple players in the real world and the virtual world quickly. Finally, HCI technology is innovating, and new attack methods are constantly being produced. From the initial USB attack to the voice attack, fingerprint cracking, etc., the development of attacks and defenses on HCI has been continuously promoted. HCI is a double-edged sword where convenience and security coexist.

\subsection{Lack of accuracy}
Although HCI is well-defined, as shown in Fig. \ref{fig:steps}, there are challenges in implementing any of these steps, and they determine the accuracy of HCI. 

\begin{itemize}
    \item \textbf{Complexity of channels}: Since the future of HCI is characterized by multimodality, the input channels will be more complex than any of the current interactions. The signals come from all parts of the human body and represent completely different kinds of senses. Therefore, the signal acquisition and integration phases need to ensure the comprehensiveness of the input and a good classification of the signal.

    \item \textbf{Ambiguity of commands}: When completing a command action, the user's planning of his or her own body movements is often deterministic, but the movements made are not exactly the same each time. In gestural interactions, for example, the user cannot get close to the same finger and wrist amplitude and dwell time each time. Although this could be simply solved by imposing more stringent input requirements on the user, it would result in a loss of naturalness of the interaction. In addition, different users may produce different interaction behaviors for the same interaction intent, which is common in voice interaction. How to make correct inferences on ambiguous natural behavior data is a challenge in the process stage.

    \item \textbf{Multiple meanings of intent}: Human emotions are complex, so there will be multiple and even contradictory expressions of the user's intention. For example, when a user expresses "tears of joy", that is, tears, accompanied by cheers and clenched fists. At this point, facial, speech, and gesture recognition may convey conflicting signals (possibly sadness, happiness, and anger, respectively), and it will be a challenge to infer the user's current state based on these complex signals.

    \item \textbf{Consistency of feedback}: It does not mean that the job is done when the machine correctly infers the user's intent and provides the appropriate feedback. How the feedback is given to the user in a more natural way will also determine the user's satisfaction with the interaction. Like the input channels, the output responsible for conveying feedback is multi-channel, including light, sound, temperature, and other forms. When these feedbacks are not coordinated and are out of sync, the experience is greatly diminished, increasing the demand for the sensitivity of the feedback channel.
\end{itemize}

\subsection{Resource consumption}
In order to create an immersive world that can rival the real world, a series of digital technologies, such as 5G networks, AI, blockchain, cloud computing, big data, IoT, VR, and AR should be developed. The IoT and the industrial Internet connect online and offline data; 5G networks provide high-speed and stable data transmission; Blockchain capitalizes data in the metauniverse to form a new trusted mechanism and collaboration models; VR and AR change the ways of HCI. The large-scale development of each technology must come at the cost of increasing consumption of energy and resources.

The Metaverse is a digital society whose computing costs are entirely new, and it involves massive data perception, transmission, computation, and rendering, which poses a huge challenge to existing network resources. The complex operation mechanism of the Metaverse requires significant network, storage, and computing resources \cite{zhang2023towards}. If it is allowed to develop blindly and disorderly, it will have a negative impact on high-quality economic and social development, energy conservation, and emission reduction. Hence, sustainable development of energy is the core constraint of the Metaverse. How to build and operate infrastructure in a green way also needs to be considered under the trend of global carbon neutrality. This requires continuous innovation and development of advanced energy infrastructure to achieve intelligent energy processing, so as to improve energy recycling efficiency. Secondly, optimize the algorithm and model of energy management, realize energy reorganization and optimization, and improve energy utilization by creating virtual resources. Currently, the relevant work includes \cite{feng2023resource,chu2023metaslicing,du2022exploring}, etc.

\section{Conclusion} \label{sec:Conclusion}

This paper focuses on the topic of interaction in Metaverse. This is the first step, i.e., the entrance to the Metaverse. We first analyze the relationship between Metaverse and HCI, which we believe are mutually reinforcing and inextricably linked. Then, we review the evolution of HCI and draw out the rules from it, i.e., the shift from machine-centered to human-centered. To this end, we provide a vision of HCI in the Metaverse era, including features and processes. In addition, while we have tried to be comprehensive in discussing the technologies needed for HCI, there may still be some that we have not introduced in this paper. We hope this article will help practitioners and researchers think about possible future interaction paradigms as they explore HCI for Metaverse. Furthermore, we have listed current advanced HCI products that are based on different technologies and are HCI companies exploring the future of HCI. Notably, this article identifies the necessary technologies that HCI relies on AI, edge computing, and others, which will determine the future experience of HCI. Finally, we analyze in detail the challenges and issues encountered by HCI, including technological constraints, a lack of accuracy, and resource consumption. We believe that in the future, interactions in the Metaverse will be easier to use, lighter, faster, more diverse, and have a better experience.

\section*{Acknowledgment}

This research was supported in part by the National Natural Science Foundation of China (Nos. 62002136 and 62272196), the Natural Science Foundation of Guangdong Province (No. 2022A1515011861), and the Young Scholar Program of Pazhou Lab (No. PZL2021KF0023). Dr. Wensheng Gan is the corresponding author of this paper.

\bibliographystyle{IEEEtran}
\bibliography{paper}

\end{document}